\documentclass[reprint,amsmath,amssymb,aps,prd]{revtex4-2}
\usepackage[colorlinks,citecolor=blue,linkcolor=blue,anchorcolor=blue,filecolor=blue,
urlcolor=blue]{hyperref}
\usepackage{graphicx}
\usepackage{ulem}
\usepackage{dcolumn}
\usepackage{bm}
\usepackage{todonotes}

\newcommand{\vis}{\mbox{\tiny vis}}

\newcommand{\fdm}{\mbox{\tiny FDM}}
\newcommand{\bdm}{\mbox{\tiny BDM}}
\newcommand{\dm}{\mbox{\tiny DM}}

\begin{document}

\title{Strange stars admixed with dark matter: equiparticle model in a two fluid approach}
\author{Isabella Marzola, Everson H. Rodrigues, Anderson F. Coelho, and Odilon Louren\c{c}o}
\affiliation{
Departamento de F\'isica e Laborat\'orio de Computa\c c\~ao Cient\'ifica Avan\c cada e Modelamento (Lab-CCAM), Instituto Tecnol\'ogico de Aeron\'autica, DCTA, 12228-900, S\~ao Jos\'e dos Campos, SP, Brazil
}

\date{\today}

\begin{abstract} 
In this work, we explore the possible scenario of strange stars admixed with fermionic or bosonic dark matter. For the description of the ``visible'' sector, we use a specific phenomenological quark model that takes into account in-medium effects for the quark masses through a suitable baryonic density dependence, in which the free parameters are chosen from the analysis of the stellar matter stability window (parametrizations presenting lower energy per baryon than iron nuclei). In the case of the dark sector, we investigate the predictions of fermionic and bosonic models. In the former, we consider a spin $1/2$ dark particle, and the latter is described by a dark scalar meson. Both models present a repulsive vector interaction, particularly important for the bosonic model since it helps avoid the collapse of the star due to the lack of degeneracy pressure. Our results point out possible stable strange stars configurations in agreement with data from the NASA NICER measurements of the pulsars PSR J0030+0451 and PSR J0740+6620, including the latest observational data, as well as the observed HESS J1731-347 mass-radius relationship. As another feature, we find stars with dark matter halo configurations for lower values of the dark particle mass.
\end{abstract}

\maketitle

\section{Introduction}
\label{sec:int}
\noindent

Stars are among the most studied astrophysical systems in the universe. However, their detailed composition, structure, and dynamics remain not fully understood by astrophysicists. For decades, these compact objects have been the subject of intense research, encompassing various areas of physics, such as thermodynamics, quantum field theory, general relativity, and nuclear reactions. The extreme conditions in the cores of stars, regions expected to be in a very high-density regime, also allow for the investigation of strongly interacting matter.

The theoretical input used to accurately describe compressed stellar matter (CSM) comes from models in which hadrons (baryons and mesons) are the fundamental degrees of freedom. Nevertheless, another branch of study considers quarks as the elementary constituents, using models that derive equations of state applied to the CSM formulation. This approach is based on the quark star hypothesis, proposed long ago in~\cite{ivanenko65,ivanenko70}, and later supported by the strange
matter hypothesis~\cite{Bodmer:1971,Witten:1984,terazawa79,terazawa89_1,terazawa89_2,terazawa90}: quark matter consisting of $u$, $d$, and $s$ quarks in beta equilibrium may have a lower energy per baryon than hadronic matter. Evidence for the presence of quark matter in the cores of massive stars has been recently discussed in Refs.~\cite{nat1,nat2}.

Both hadronic and phenomenological quark models, developed for the aforementioned purpose, can be tested against a vast number of recent astrophysical observations. These new measurements have emerged primarily due to:~(i) the detection of gravitational waves from a binary system, along with their respective electromagnetic counterparts~\cite{ligo-em}, as observed by the LIGO and Virgo Collaboration~\cite{Abbott_2017,Abbott_2018,Abbott_2020-2};~(ii) data from massive millisecond pulsars PSR J0030+0451~\cite{Miller_2019,Riley_2019} and PSR J0740+6620~\cite{Fonseca_2021,Riley_2021,Miller_2021,Salmi_2024,Dittmann_2024}, provided by NASA's Neutron Star Interior Composition Explorer~(NICER), an X-ray telescope aboard the International Space Station~\cite{nat4}; and~(iii) the first simultaneous measurements of mass, radius, and surface temperature of the compact star HESS J1731-347~\cite{nat3}.

Recently, many researchers have dedicated efforts to better understanding the physics of these objects, which may also contain another mysterious component known as dark matter (DM)~\cite{revdm1,revdm2,revdm3}. The main evidence for the existence of non-luminous matter comes from the analysis of galaxy rotation curves and gravitational lensing methods~\cite{lensing1,lensing2}, which indicate that the visible mass of galaxy clusters accounts for only about 10\% to 20\%, with the remaining mass attributed to DM. Measurements of the anisotropy of the cosmic microwave background also suggest the presence of DM~\cite{cmb1,cmb2,cmb3,cmb4,cmb5,cmb6}. It is widely accepted that the universe, as we know it today, is composed of approximately 27\% DM, 68\% dark energy, and 5\% ordinary matter.

In this work, we investigate the possibility of the existence of strange stars admixed with fermionic and bosonic dark matter, in accordance with the astrophysical constraints mentioned earlier. To this end, we employ a specific effective quark model in the ``visible'' sector, namely, the equiparticle~(EQP) model in which the constituent quark masses ($m_{u,d,s}$) are medium dependent, specifically density-dependent functions in the case of stellar matter at $T=0$. The effects of temperature on $m_{u,d,s}$ have been explored in~\cite{adamu,Wen:2005,debora24,xia22}, while in~\cite{adamu2}, modified density dependencies of these quantities have been proposed. The thermodynamical inconsistency of the original EQP model was resolved in~\cite{Xia:2014} by establishing a suitable connection between the Fermi momentum of the quark and the corresponding effective chemical potential. In~\cite{Marzola:2023}, some of us further improved this model to incorporate deconfinement effects by introducing the traced Polyakov loop as the order parameter for the first-order confinement/deconfinement phase transition. This revised formulation was named the PEQM model.

Here, we adopt the two-fluid approach, which considers only gravitational interaction between DM and quark matter. The calculations and results obtained in this investigation, all expressed in natural units ($\hbar = c = G = 1$), are organized as follows. In Sec.~\ref{sec:eqpm}, we present the main quantities related to the effective EQP model and establish the conditions for the physical stability window. In Sec.~\ref{sec:dm}, we detail the main equations of state used to describe the dark sector in our analysis, including fermionic and bosonic DM models. The results concerning strange stars derived from this formalism are presented in Sec.~\ref{sec:strange}. Finally, our concluding remarks are provided in Sec.~\ref{sec:summary}.

\section{Equiparticle model}
\label{sec:eqpm}

\subsection{Formalism}

The EQP model was proposed in~\cite{Xia:2014} as a baryonic density~($\rho_{b}$) and temperature~($T$) dependent model. However, here we focus on the zero temperature regime, as done in~\cite{Backes:2020} and~\cite{Marzola:2023}, for the study of cold quark stars. For this reason, the formalism of the model will be presented for this case. For an infinitely large system composed of up~($u$), down~($d$), and strange ($s$) quarks, the equations of state are obtained from fundamental thermodynamical relations. For the pressure we have
\begin{equation}
P = -\Omega_{0}+\rho_b \frac{\partial m_I}{\partial \rho_b} \frac{\partial \Omega_{0}}{\partial m_I},
\label{eq:pressure}
\end{equation}
where $m_I$ is the density-dependent quark mass term. The energy density reads
\begin{equation}
\mathcal{E} = \Omega_0 - \sum_i \mu_i^* \frac{\partial\Omega_0}{\partial\mu_i^*},
\label{eq:energy-density}
\end{equation}
in which $\mu_i^*$ is the effective chemical potential that is introduced through the Fermi momentum ($k_{Fi} = \sqrt{\mu_i^{*2} - m_i^2}$) in order to ensure the thermodynamic consistency, as explained in~\cite{Peng:2008}, and $m_i$ is the constituent quark mass ($i=u,d,s$). Furthermore, it is possible to obtain the real chemical potentials of the system by the relationship
\begin{equation}
    \mu_i = \mu_i^* + \frac{1}{3}\frac{\partial m_I}{\partial \rho_b} \frac{\partial \Omega_0}{\partial m_I}.
    \label{eq:real-effective-chemical-potentials}
\end{equation}
For all equations presented, the quantity identified as $\Omega_0$, that is responsible for the free system particle contribution, is given by~\cite{Peng:2000},
\begin{equation}
\begin{split}  
&\Omega_0 =\\
&- \sum_i \frac{\gamma}{24 \pi^2} \left[ \mu_i^*k_{Fi} \left( k_{Fi}^2 - \frac{3}{2}m_i^2 \right) + \frac{3}{2} m_i^4 \ln \frac{\mu_i^* + k_{Fi}}{m_i} \right],\\
\end{split}
\label{eq:thermodynamic-potential-free-system}
\end{equation}
with $\gamma=6=3\mbox{ (color)}\times 2\mbox{ (spin)}$ being the degeneracy factor. Concerning the grand-canonical thermodynamic potential, it can be obtained as $\Omega = -P$. 

In order to fully describe the model, it is also important to introduce the relations between quark densities ($\rho_i$) and other quantities, such as the Fermi momentum and the baryonic density that are, respectively, given by
\begin{equation}
\rho_i = \frac{\gamma k_{Fi}^{3}}{6\pi^2},
\label{eq:particle-density}
\end{equation}
and
\begin{equation}
\rho_b = \frac{1}{3} \sum_i \rho_i.
\label{eq:baryon-number-density}
\end{equation}

With regard to the thermodynamical consistency of the model, one can verify this characteristic by different means. The first verification concerns to the condition
\begin{equation}
\Delta = P - \rho_b^{2}\frac{d}{d\rho_b}\left(\frac{\mathcal{E}}{\rho_b}\right) = 0.
\end{equation}
Any consistent thermodynamic model must ensure that $\Delta$ = 0 or, in other words, the pressure at the minimum energy per baryon must be zero. A second verification that can be done is through the Euler relation given by
\begin{equation}
\begin{split}
 P + \mathcal{E} &= \sum_{i}\mu_{i}\rho_{i}.
 \label{eq:consist-verification2}
\end{split}
\end{equation}

Finally, the third option is to analyze the derivative of $-\Omega_0$ with respect to $\mu_i^*$ as presented by
\begin{equation}
\rho_i = -\frac{\partial\Omega_0}{\partial\mu_i^*},
\label{eq:consist-verification}
\end{equation}
where the comparison with the particle number density in Eq.~\eqref{eq:particle-density} results in an equivalence. These procedures leads us to conclude that all equations of the model are consistent with the ones from fundamental thermodynamics, regardless the choice of the quark mass scaling, such as the ones from~\cite{Fowler:1981,Chakra:1991,Peng:2000,Xiaoping:2004}.

Following up on this, for the EQP model the quark masses are constructed as being density-dependent as follows:
\begin{equation}
m_{i} = m_{i0} + m_I = m_{i0} + \frac{D}{\rho_b^{1/3}} + C\rho_b^{1/3},
\label{eq:masses}
\end{equation}
where $m_{i0}$ is the current mass of the quark $i$. The interaction effects between the quarks are contained in the parameters $C$ and $D$, which are chosen from a stability window which will be described later. When working with compact objects, there is an interest in achieving higher stellar masses to satisfy the observational constraints. For this reason, the parameter $C$ plays an important role highlighting the necessity of incorporating this parameter into the mass scaling derived from~\cite{Xia:2014} and presented in Eq.~\eqref{eq:masses}.

\subsection{Stellar matter}

Given the interest in studying compact objects, it is necessary to establish the conditions for the matter present in its interior. At zero temperature and high-density regime, stellar matter in pure quark or hybrid stars~\cite{Schertler:1999, Ranea:2019, Hanauske:2001, Menezes:2006, Lenzi:2010, Drago:2020, Pereira:2016, Ferreira:2020} can be described by effective quark models. 

The most powerful mechanism of neutrino emission in stars is the Urca process~\cite{Lattimer:1991,YAKOVLEV:20011}, responsible for their cooling since in this process stars lose a significant amount of thermal energy. The rate of neutrino emission influences how quickly the star cools down from its initial high temperature. After neutrinos are emitted from the star, its internal conditions undergo some changes, where the beta equilibrium takes place. This condition is given by  $\mu_u + \mu_e = \mu_d = \mu_s$ in quark stars, which are the chemical potentials of the particles involved. In terms of the previously discussed model, this relation can be written as $\mu_u^* + \mu_e = \mu_d^* = \mu_s^*$, where the chemical potentials of the quarks are now the effective ones. The electron ($e$) does not have an effective chemical potential because its mass is constant. This equilibrium condition ensures that the star remains stable and maintains its composition over cosmic timescales. Furthermore, another condition has to be satisfied in compact stars, namely, the charge neutrality condition, given here by $\frac{2}{3}\rho_u - \frac{1}{3}\rho_d - \frac{1}{3}\rho_s - \rho_e = 0$. These conditions exist due to the presence of leptons (electrons in this case), where weak interactions happen such as $d, s \leftrightarrow u + e + \bar\nu_e$. Because of the electrons' presence, stellar matter is described by the sum of quarks and electrons' contribution. Considering the (massless) electrons contribution, the total energy density and pressure of the part of the system composed of ``visible'' matter are
\begin{equation}
\mathcal{E}_{\vis} = \mathcal{E} + \frac{\mu_e^4}{4\pi^2}
\label{eq:det}
\end{equation}
and
\begin{equation} 
P_{\vis} = P + \frac{\mu_e^4}{12\pi^2},
\label{eq:presst}
\end{equation}
respectively, with $\mu_e=(3\pi^2\rho_e)^{1/3}$.

\subsection{Stability window}

For the discussion of strange quark matter~(SQM) in the context of quark stars, one has to consider the Bodmer-Witten hypothesis~\cite{Bodmer:1971,Witten:1984}. It suggests that under certain conditions, such as at high densities or pressures, SQM may be the true ground state of matter rather than normal nuclear matter. Therefore, in order to establish the stability of SQM we need to investigate it against nuclear matter, which can be done by evaluating the minimum of the energy per baryon, $(E/A)_{\mbox{\tiny min}}$. This evaluation is then classified as follows: (i) SQM is stable when the minimum of the energy per baryon is lower than the binding energy of $^{56}$Fe (because it is the heaviest stable nucleus found in nature), i.e., $(E/A)_{\mbox{\tiny min}}\leqslant 930$~MeV; (ii) SQM is considered to be meta-stable when the minimum of the energy per baryon is higher than 930 MeV and lower than 939 MeV, which is the nucleon mass, i.e., $930\mbox{ MeV}<(E/A)_{\mbox{\tiny min}}\leqslant 939$~MeV; (iii) SQM is unstable only if the minimum of the energy per baryon is higher than 939 MeV, i.e, $(E/A)_{\mbox{\tiny min}}> 939$~MeV. In the case of stellar matter, we have $E/A = \mathcal{E}_{\vis}/\rho_b$.

\begin{figure}[!htb]
\centering
\includegraphics[scale=0.21,trim=0 2cm 0 3cm,clip=true]{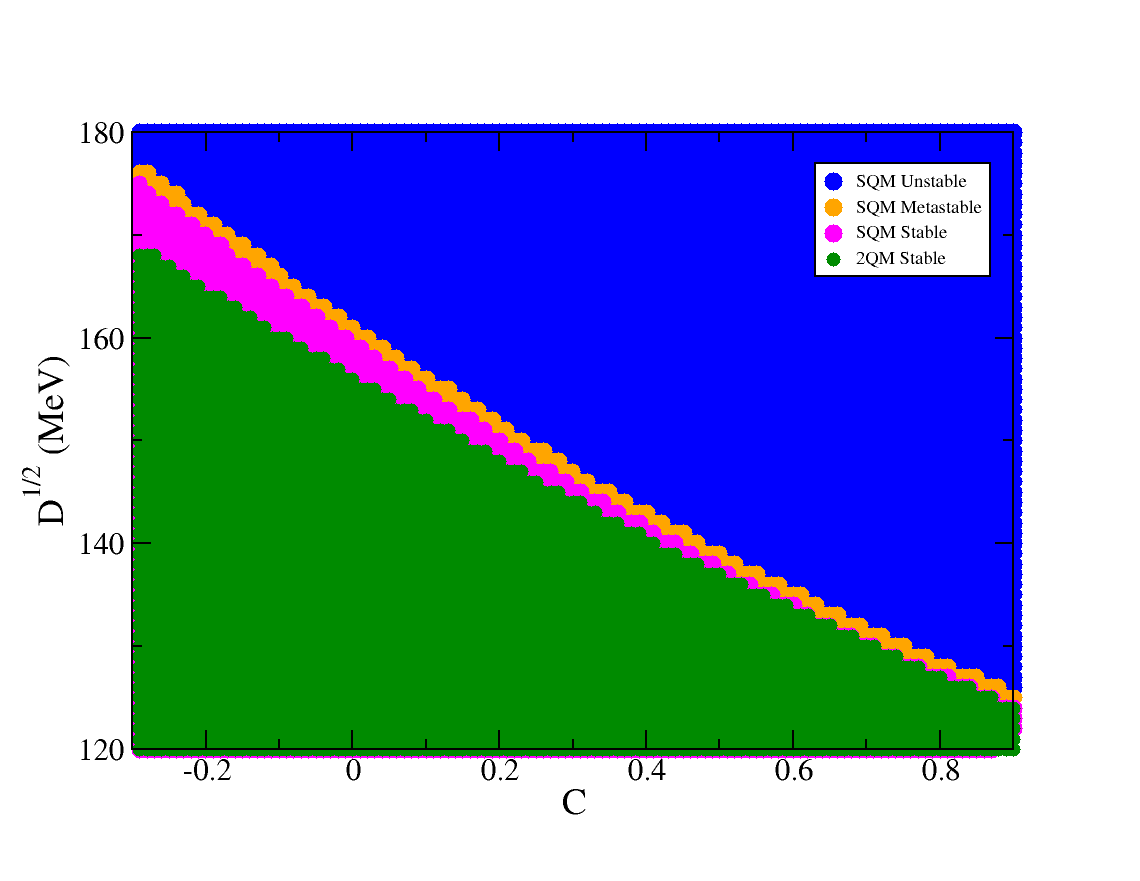}
\caption{Stability window for stellar matter predicted by EQP model.}
\label{stability-window-stellar-matter}
\end{figure}

With the minimum of the energy per baryon in hand, it is possible to construct what is called the stability window for a large set of values for the $C$ and $\sqrt{D}$ parameters of the EQM model. This analysis is depicted in Fig.~\ref{stability-window-stellar-matter}.
In the figure, we observe different regions of stability for SQM. The pink one corresponds to the absolutely stable state of SQM, which is our area of interest. Below this area, we find the green region where two-flavor quark matter~(2QM) is stable and where the SQM is forbidden to happen. In the orange region, the meta-stable state of SQM is expected to occur, and in blue we have the unstable SQM. It is important to mention that this analysis takes into consideration the data provided by PDG~\cite{pdg22} regarding the ranges for the current quark masses, $m_{u0} = 2.16^{+0.49}_{-0.26}$, $m_{d0} = 4.67^{+0.48}_{-0.17}$ MeV and $m_{s0} = 93.4^{+8.6}_{-3.4}$ MeV. For stellar matter the chosen values are $m_{u0} = 2.16$~MeV~, $m_{d0} = 5.15$~MeV, and $m_{s0} = 90$~MeV. For the pair $(C,D)$ we take $C = 0.81$ and $D^{1/2} = 127$~MeV, since it produces more massive stars when only quark matter is considered.

\begin{figure}[!htb]
\includegraphics[scale=0.29]{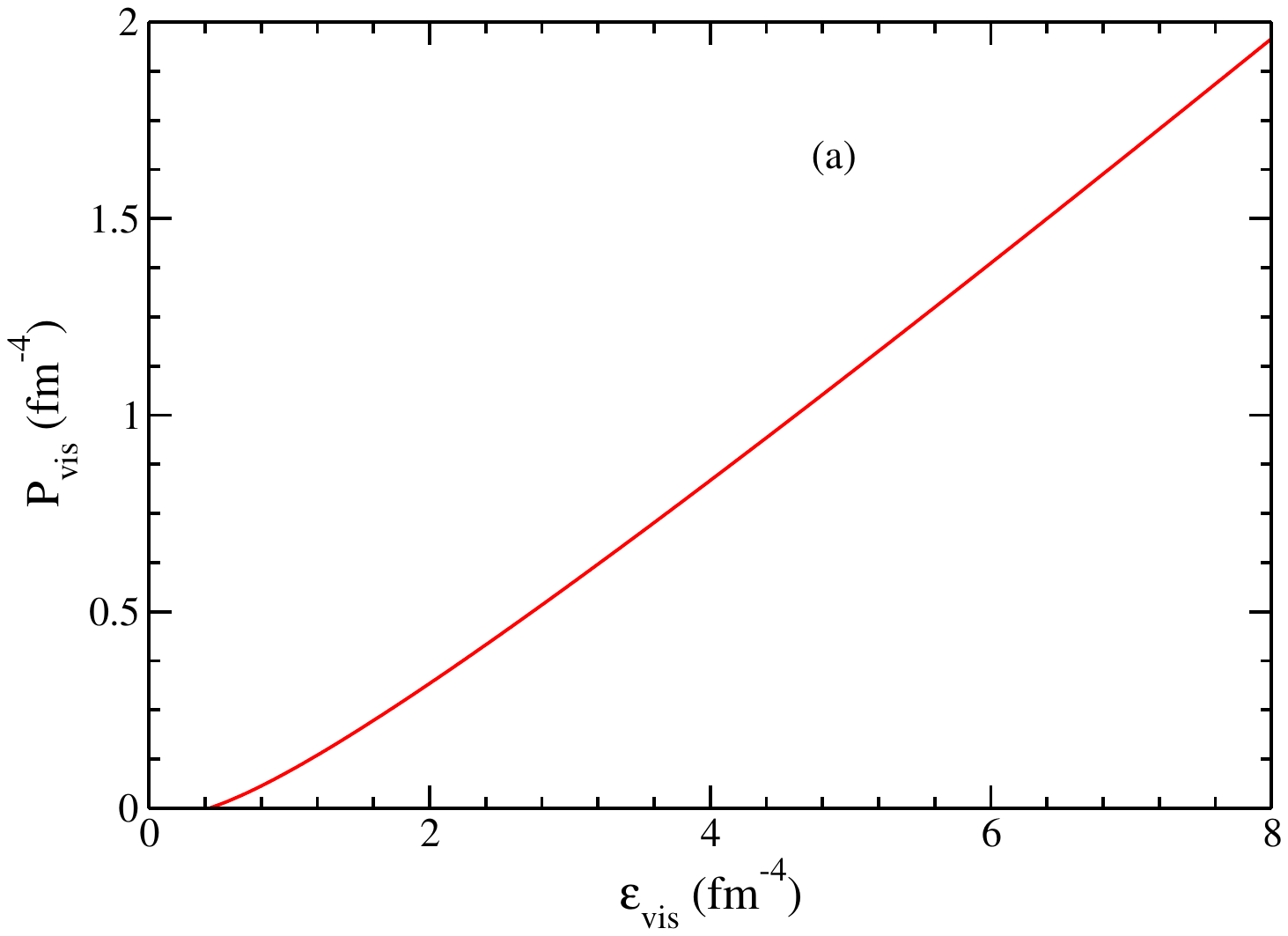}\vspace{-0.8cm}
\includegraphics[scale=0.29]{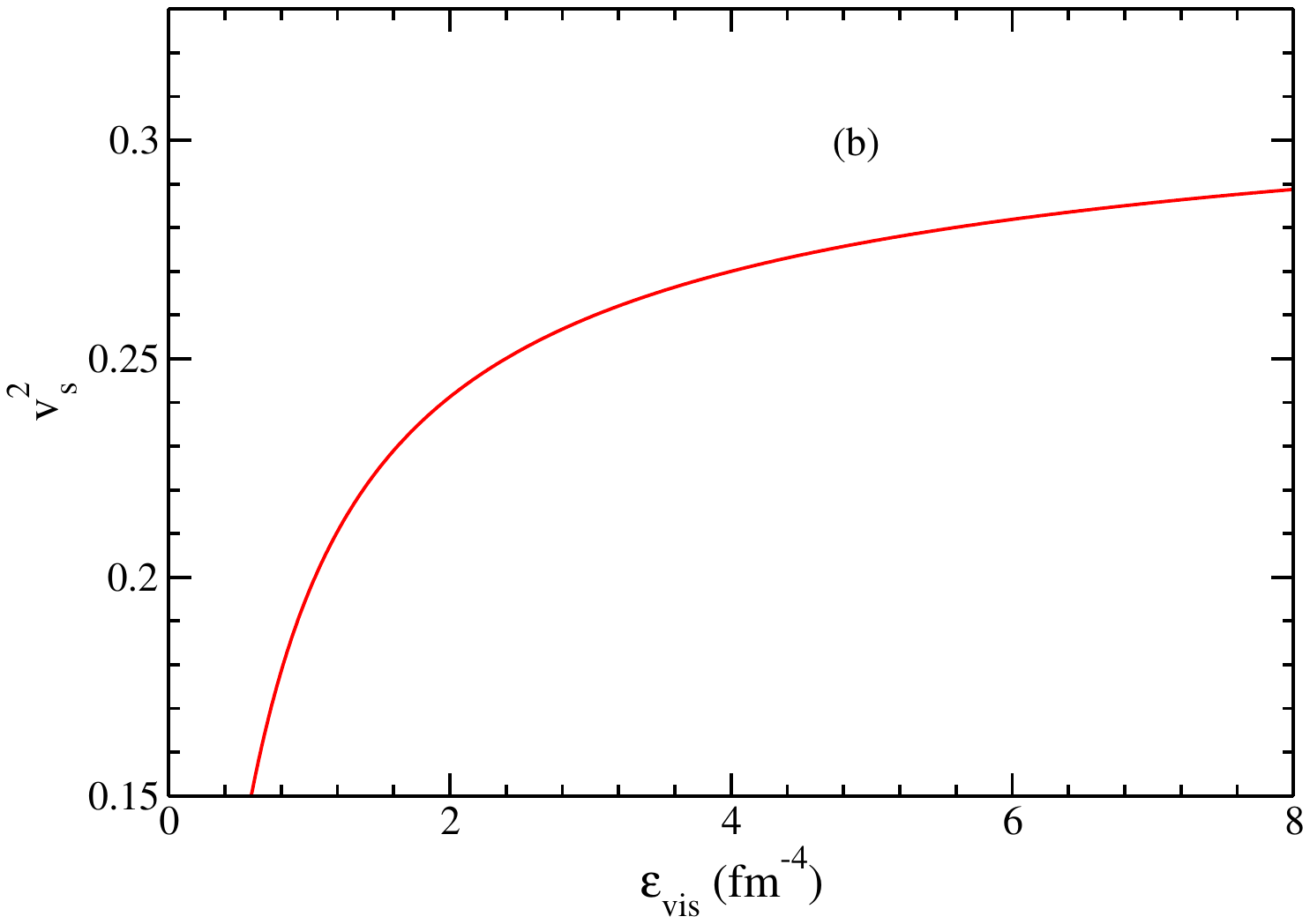}
\caption{Panel (a): dark matter pressure as a function of its respective energy density. Panel (b): dark matter sound speed. Fermionic and bosonic cases are represented by full and dashed lines, respectively.}
\label{fig:eos-eqpm}
\end{figure}

For the sake of completeness, we show in Fig.~\ref{fig:eos-eqpm} the equation of state of the EQPM model, along with its respective sound speed, for the stellar (beta equilibrated) matter case. 
Notice that Fig.~\ref{fig:eos-eqpm}{\color{blue}b} confirms that the model satisfies causality, i.e., it presents no superluminal states for the energy density range used in the work.

\subsection{Polyakov equiparticle model}

The Polyakov equiparticle (PEQP) model was proposed in~\cite{Marzola:2022a,Marzola:2023} to study the confinement/deconfinement phase transition, possible to happen inside strange stars. In that article, some of us perform a few modifications in the original EQP model, which will be briefly described in this section. In this improved model, gluons and their corresponding strong interaction dynamics are phenomenologically described by the traced Polyakov loop, $\Phi$. When $\Phi=0$, the quarks are confined, and with $\Phi\rightarrow 1$ they achieve the deconfined phase. In the NJL model~\cite{njl1,njl2,njl3,njl4,njl5,njl6}, this quantity was introduced, at finite temperature~\cite{Fukushima:2004}, to generate the so-called Polyakov-Nambu-Jona-Lasinio (PNJL) model~\cite{Ratti:2006, Robner:2007, Ratti:2007, Fukushima:2008, Dexheimer:2010, Dexheimer:2021,Robner:2009}. At first, the PNJL model loses its characteristics of describing the confinement/deconfinement phase transition at $T=0$, but in~\cite{sagun} the authors were able to solve this problem by taking the $b_2(T)$ function, present in the Polyakov potential, and modifying it to become a function that is also dependent on the quark density. Later on, the PNJL model suffered another improvement done in~\cite{Mattos:2019,Mattos:2021,Mattos:prd}, where 
the motivation was to ensure that the coupling constants would vanish at the deconfinement phase, i.e. at $\Phi=1$. 

All these observations concerning the improvement of the original NJL model were also implemented in the EQP model. More specifically, the parameters $C$ and $D$ were modified as follows, $C\rightarrow C'(C,\Phi)=C(1-\Phi^2)$, and $D\rightarrow D'(D,\Phi)=D(1-\Phi^2)$, which lead us to a new equation for the quark masses,
\begin{align}
m_{i}(\rho_b,\Phi) = m_{i0} + \frac{D(1-\Phi^2)}{\rho_b^{1/3}} + C(1-\Phi^2)\rho_{b}^{1/3},
\label{eq:masses-peqpm}
\end{align}
i.e., now $m_i$ is also a function of $\rho_b$ and $\Phi$ as well. For the equations of state, energy density and pressure respectively, we have that $\mathcal{E}_{\mbox{\tiny PEQP}} = \mathcal{E}_{\mbox{\tiny EQP}}+\mathcal{U}_{0}(\Phi)$ and $P_{\mbox{\tiny PEQP}}=P_{\mbox{\tiny EQP}}-\mathcal{U}_{0}(\Phi)$ where $\mathcal{U}_0(\Phi)$ is included to ensure $\Phi\ne 0$ solutions and also to limit $\Phi$ in the physical range of $0\leqslant\Phi\leqslant 1$, according to the findings of~\cite{Mattos:2019,Mattos:2021} (we also consider that $\Phi=\Phi^*$). This term is given as $\mathcal{U}_{0}(\Phi) = a_{3}T_{0}^{4}\mbox{ln}(1 - 6\Phi^{2} + 8\Phi^{3} - 3\Phi^{4})$, with $a_{3}$ being a dimensionless free parameter, and $T_0 =$ the transition temperature for the pure gauge system. Notice that if we take $a_{3} = 0$, the original EQP model is restored, since this leads to $U_{0} = 0$ and, consequently, to $\Phi=0$, $C'=C$, and $D'=D$. The PEQM model was only introduced here to acknowledge that this model is being studied to explain different physical phenomena in quark matter, but our choice in the present work is to study dark matter inside strange stars by using the original EQP model, i.e., by taking, for the sake of simplicity, $a_3=0$.

\section{Dark matter models}
\label{sec:dm}

With regard to the dark sector of our approach, we choose to investigate the capability of the DM model in satisfying the recent astrophysical constraints when admixed to the previously described quark matter, by implementing a fermionic and a bosonic model. Concerning the first one, we adopt, as a starting point, a Lagrangian density presenting a kinetic term (Dirac Lagrangian density) of a single fermionic component, along with a vector meson coupled to the Dirac spinor. The full expression is given by~\cite{qian14,tuhin22,tuhin24}
\begin{align}
\mathcal{L}_{\fdm} &= \bar{\chi}\left[ \gamma_{\mu} (i\partial^{\mu} - g_VV^{\mu}) - m_{\chi} \right]\chi - \frac{1}{4}F_{\mu\nu}F^{\mu \nu}
\nonumber\\
&+ \frac{1}{2}m^{2}_VV_{\mu}V^{\mu},
\label{eq:lagdendm}
\end{align}
where the strength tensor of the vector meson is $F_{\mu\nu} = \partial_\mu V_\nu - \partial_\nu V_\mu$. The dark fermion mass is $m_\chi$, and $m_V$ is the mass of the dark vector meson. We do not include here the dark scalar meson, as done in~\cite{qian14}, in order to simplify the model and avoid one more free parameter to be fixed. Besides that, the main effect of increasing the final compact star mass is due to repulsion interactions. Therefore, the inclusion of a scalar particle would add attraction to the system and, consequently, a decreasing of the repulsion as a net feature. 

In a similar way as done in relativistic hadronic models that present the same mathematical formulation for the Lagrangian density, we employ here the mean-field approach in order to compute energy density and pressure of the dark sector of the system. These expressions read
\begin{align}
\mathcal{E}_{\fdm} &= \frac{1}{\pi^2}\int_0^{k_{F\chi}}\hspace{-0.2cm}dk\,k^2(k^2 + m_\chi^2)^{1/2}
+\frac{1}{2}C_V^2\rho_\chi^2,
\label{eq:defer}
\\
P_{\fdm} &= \frac{1}{3\pi^2}\int_0^{k_{F\chi}}\hspace{-0.2cm}dk\,\frac{k^4}{(k^2 + m_\chi^2)^{1/2}}
+\frac{1}{2}C_V^2\rho_\chi^2,
\label{eq:pfer}
\end{align}
with $C_V=g_V/m_V$ and $\rho_\chi=k_{F\chi}^3/(3\pi^2)$. The Fermi momentum of the dark particle is $k_{F\chi}$. Fermionic self-interacting dark matter models coupled to hadronic matter were used in the description of neutron stars with DM content, as the reader can find in~\cite{milva24}, for instance. 

Despite efforts dedicated to the direct and indirect detection methods for dark matter in the last years~\cite{detections}, the true nature of DM particles is still not well established. This is the origin of the many possible candidates studied over the years, as the weakly interacting massive particles, gravitinos, axinos, axions, sterile neutrinos, WIMPzillas, supersymmetric Q-balls, and mirror matter~\cite{cand1,cand2}. Among these options, both fermionic and bosonic dark particles are considered. In order to take into account the latter ones, we assume the bosonic asymmetric dark matter model proposed in~\cite{nelson19} and explored in detail in~\cite{watts23}. After neglecting interactions between the dark boson with visible matter, and approximating the spacetime to be flat, the Lagrangian density for the model reads
\begin{align}
\mathcal{L}_{\bdm} = &-\sqrt{-g}\left(D^*_\mu\sigma^*D^\mu\sigma - m^2_\sigma\sigma^*\sigma - \frac{1}{2}m_\phi^2\phi_\mu\phi^\mu\right.
\nonumber\\
&- \left. \frac{1}{4}Z_{\mu\nu}Z^{\mu \nu}\right),
\label{eq:lagbos}
\end{align}
where $D_\mu = \partial_\mu + ig_\sigma\phi_\mu$, $g_\sigma$ is the interaction
strength of the dark scalar complex field $\sigma$ with the dark vector field $\phi^\mu$, $Z_{\mu\nu} = \partial_\mu \phi_\nu - \partial_\nu \phi_\mu$, and $g$ is the determinant of the metric. The masses of the dark scalar and dark vector fields are $m_\sigma$ and $m_\phi$, respectively. Usual quantum field theory techniques, along with the mean-field approximation, lead to the following energy-momentum tensor
\begin{align}
T_{\mu\nu} &= 2D_\mu^*\sigma^*D_\nu\sigma - g_{\mu\nu}\left(D_\rho^*\sigma^*D^\rho\sigma + m_\sigma^2\sigma^*\sigma \right)
\nonumber\\
&+m^2_\sigma\left(\phi_\mu\phi_\nu-\frac{1}{2}g_{\mu\nu}\phi_{\rho}\phi^{\rho}\right),
\end{align}
that is used to find the equations of state for the bosonic DM model given by
\begin{align}
\mathcal{E}_{\bdm} &= m_{\sigma}\rho_{\sigma} 
+ \frac{1}{2}C^2_{\sigma\phi}\rho^{2}_{\sigma},
\label{eq:debos}
\\
P_{\bdm} &= \frac{1}{2}C^2_{\sigma\phi}\rho^{2}_{\sigma},
\label{eq:pbos}
\end{align}
with $C_{\sigma\phi}=g_\sigma/m_{\phi}$. The DM density $\rho_\sigma$ relates to the zero component of the vector field $\phi_0$, and with the scalar field and its conjugate through $\phi_0=(g_\sigma/m_\phi^2)\rho_\sigma$, and $\rho_\sigma=2m_\sigma\sigma^*\sigma$, respectively. Detailed derivation of these expressions can be found in~\cite{watts23}. We also address the reader to~\cite{khlopov85} for a deeper study of scalar fields, in general, in the gravitational context. Other similar bosonic DM models coupled to luminous matter, studying neutron stars in particular, were analyzed in~\cite{davood1,davood2}.

In the analysis of bosonic DM inside compact stars, it is important to emphasize that Black holes can be formed when the Chandrasekhar limit is achieved~\cite{samuel12}. In the case of fermionic matter in the ground state, the Pauli exclusion principle helps balance gravity since degeneracy pressure is present. For bosonic systems, on the other hand, the lack of Fermi pressure sets this limit to a smaller value. However, it can be increased due to repulsive interactions like the one included in Eq.~(\ref{eq:lagbos}). In order to satisfy this condition, we restrict the values of $m_\sigma$ and $C_{\sigma\phi}$ to 
$10^{-2}~\mbox{MeV} \leqslant m_\sigma \leqslant 10^8~\mbox{MeV}$ and 
$10^{-2}~\mbox{MeV}^{-1} \leqslant C_{\sigma\phi} \leqslant 10^3~\mbox{MeV}^{-1}$, respectively, in agreement to the findings of~\cite{watts23}.

\section{Strange stars admixed with dark matter}
\label{sec:strange}

There are at least two main approaches used to describe strongly interacting matter admixed with dark matter at the mean-field level. One of them is based on the Higgs-portal models~\cite{higgsportal1,higgsportal2}, where DM is assumed to interact with ordinary particles through the exchange of the Higgs boson, with the whole system (DM + ``visible sector'') described by a single Lagrangian density. It is shown that in hadronic models coupled to DM, this particular procedure leads to a simplification of the final equations of state, leading to a vanishing coupling between the distinct sectors (see~\cite{lourenco22} for instance). From the experimental point of view, such a result is corroborated by direct detection experiments~\cite{LUX:2017ree,XENON:2018voc,PandaX-II:2017hlx,Billard:2021uyg,Schumann:2019eaa}, which only give upper bounds on the elastic scattering cross-section. Since DM particles and ordinary matter seem to be not interacting, it is reasonable to treat them as a two-fluid system in the interior of compact stars (quark, neutron, or hybrid stars, for instance); i.e., both sectors are coupled to each other only through gravity.

For the construction of strange stars through the EQP model admixed with fermionic or bosonic dark matter, all of them previously discussed, we consider the interfluid interaction uniquely coming from gravitational effects, as discussed above. For this purpose, the two-fluid formalism is used here. In this approach, each fluid satisfies the conservation of energy-momentum separately, which is equivalent to having $P(r)=P_{\vis}(r)+P_{\dm}(r)$ and $\mathcal{E}(r)=\mathcal{E}_{\vis}(r)+\mathcal{E}_{\dm}(r)$, with $r$ being the radial coordinate from the center of the star. This consideration leads to the following differential TOV equations to be solved, 
\begin{align}
\frac{dP_{\vis}(r)}{dr} &= -\frac{\left[\mathcal{E}_{\vis}(r) + P_{\vis}(r)\right]\left[m(r) + 4\pi r^3 P(r)\right]}{r\left[r - 2m(r)\right]},
\label{eq:pvis}
\\
\frac{dP_{\dm}(r)}{dr} &= -\frac{\left[\mathcal{E}_{\dm}(r) + P_{\dm}(r)\right]\left[m(r) + 4\pi r^3 P(r)\right]}{r\left[r - 2m(r)\right]},
\label{eq:pdm}
\\
\dfrac{dm_{\vis}(r)}{dr} &= 4\pi r^2\mathcal{E}_{\vis}(r),
\label{eq:mvis}
\\
\dfrac{dm_{\dm}(r)}{dr} &= 4\pi r^2\mathcal{E}_{\dm}(r),
\label{eq:mdm}
\end{align}
where $m(r) = m_{\vis}(r) + m_{\dm}(r)$ is the total mass contained
in the sphere of radius $r$. The visible matter mass is $m_{\vis}(r)$, and the dark matter mass is $m_{\dm}(r)$. In these equations, one has DM = FDM (BDM) for fermionic (bosonic) dark matter. This set of coupled equations can be obtained from the stationary condition of the star mass. For a detailed derivation, we address the reader to the appendix of~\cite{qian14}.

Technically, the procedure adopted to solve Eqs.~\eqref{eq:pvis}-\eqref{eq:mdm} is the following: first, we define the four initial conditions as $m_{\vis}(0)=m_{\dm}(0)=0$, $P_{\vis}(0)=P^c_{\vis}$, and $P_{\dm}(0)=P^c_{\dm}$, where $P^c_{\vis}$ and $P^c_{\dm}$ are the central pressures related to visible and dark matter, respectively, given by the equations of state presented in the previous sections. Then, for each set of initial conditions we use the fourth-order Runge–Kutta method in order to obtain pressures and masses as functions of $r$. The radii $R_{\vis}$ and $R_{\dm}$ are defined as being the quantities that lead to $P_{\vis}(R_{\vis})/P^c_{\vis}=0$ and $P_{\dm}(R_{\dm})/P^c_{\dm}=0$, within a certain tolerance. These radii are also useful to determine $M_{\vis}\equiv m_{\vis}(R_{\vis})$, and $M_{\dm}\equiv m_{\dm}(R_{\dm})$. Therefore, the total mass of the respective star is $M=M_{\vis}+M_{\dm}$, and its radius is $R=R_{\vis}$ if $R_{\vis}>R_{\dm}$, or $R=R_{\dm}$ if $R_{\dm}>R_{\vis}$. This latter case identifies dark matter halo configurations for the star of mass $M$ and radius $R$. This mechanism is repeated for each point of the input equations of state, with the final collected pairs $(M,R)$ giving rise to the mass-radius diagram. 

\begin{figure}[!htb]
\includegraphics[scale=0.29]{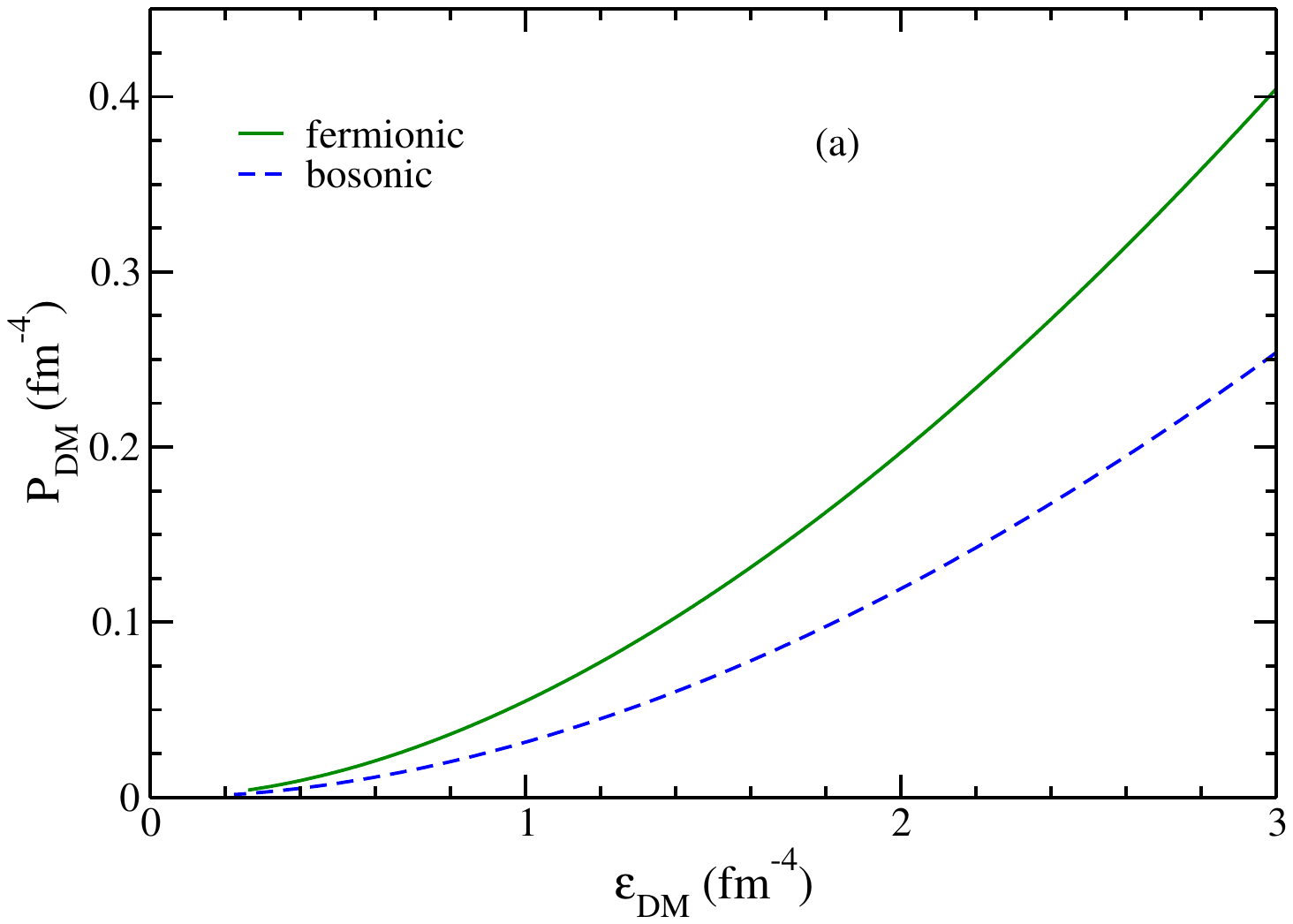}\vspace{-0.8cm}
\includegraphics[scale=0.29]{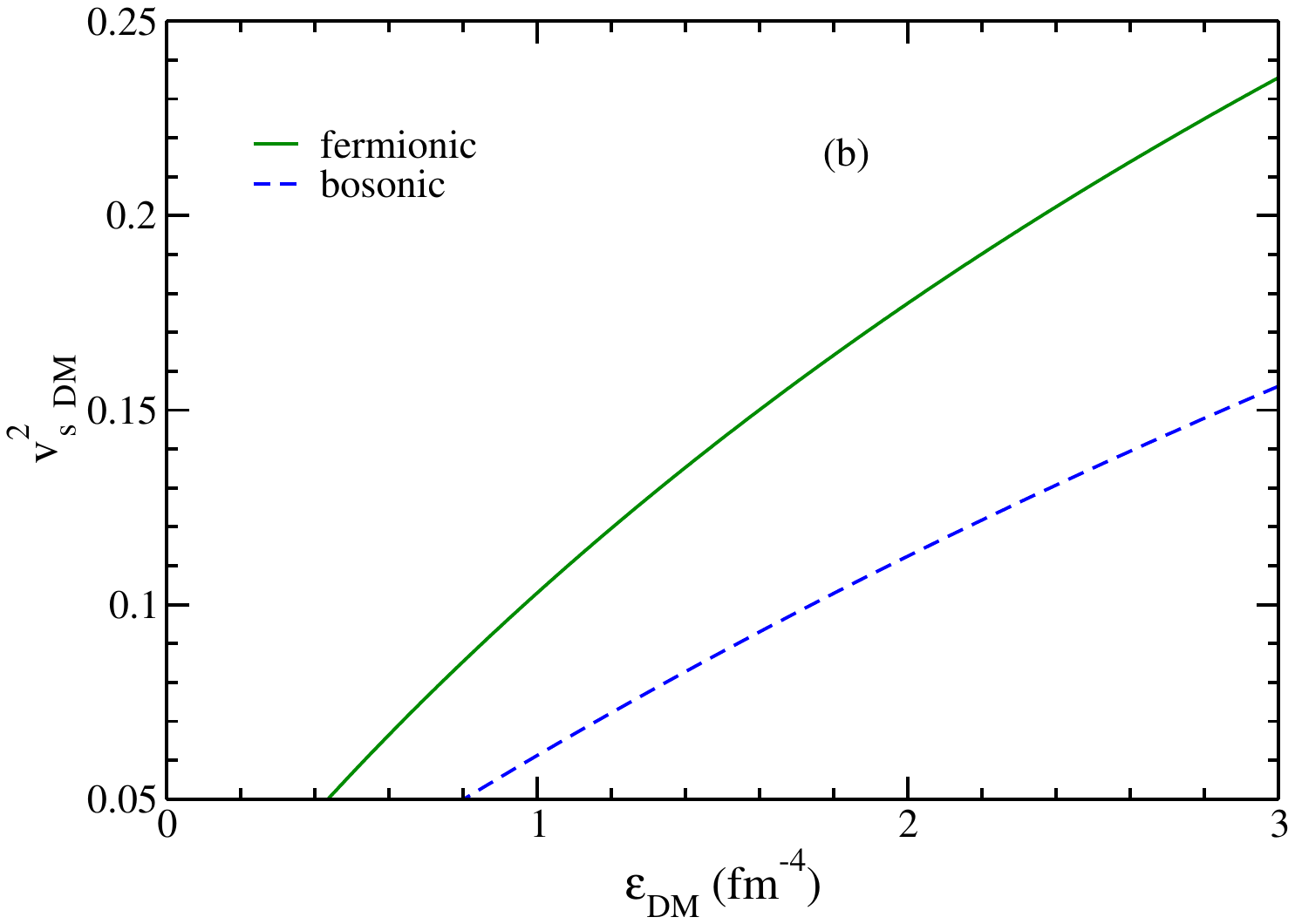}
\caption{Panel (a): dark matter pressure as a function of its respective energy density. Panel (b): dark matter sound speed. Fermionic and bosonic cases are represented by full and dashed lines, respectively.}
\label{fig:eosdm-f}
\end{figure}

Here we impose that every star presents a certain fraction of dark matter in its composition, i.e., a fixed value of $F_{\dm} = M_{\dm}/M$.
In order to satisfy this restriction, we proceed as follows. First, one needs to construct all the inputs for the TOV equations, namely $P^c_{\vis}$, $\mathcal{E}^c_{\vis}$, $P^c_{\dm}$ and $\mathcal{E}^c_{\dm}$ (although not directly used as initial conditions, the energy densities have also to be furnished since their relationships with the pressures are used to replace $\mathcal{E}$'s by $P$'s in the TOV equations). The visible matter equation of state~(EOS) is straightforward, directly evaluated from Eqs.~\eqref{eq:det}-\eqref{eq:presst}. For both fermionic and bosonic DM models, we connect them with the EQP model by imposing that $\mathcal{E}_{\dm}=f\mathcal{E}_{\vis}$ where the energy density fraction, $f$, is a constant~\cite{qian14}. In the case of the bosonic DM model, this leads to an analytical expression for the dark boson density,
\begin{align}
\rho_\sigma = \frac{m_\sigma}{C_{\sigma\phi}^2}
\left( \sqrt{1 + \frac{2C_{\sigma\phi}^2}{m_\sigma^2}f\mathcal{E}_{\vis}} - 1  \right),
\end{align}
and, consequently, to a direct relation between dark pressure and quark energy density given by
\begin{align}
P_{\bdm} &= \frac{m_\sigma^2}{2C_{\sigma\phi}^2}
\left( \sqrt{1 + \frac{2C_{\sigma\phi}^2}{m_\sigma^2}f\mathcal{E}_{\vis}} - 1  \right)^2.
\end{align}
For the fermionic dark matter case, on the other hand, there are no simple expressions relating the equations of states of quark and dark matter. The free parameters of the dark sector used here are $m_{\sigma} = 15$~GeV~\cite{watts23}, $C_{\sigma\phi}=0.1$~MeV$^{-1}$~\cite{watts23}, $m_\chi=1.9$~GeV, and $C_V=3.26$~fm. The last two values, inside the ranges $0.5~\mbox{GeV} \leqslant m_\chi \leqslant 4.5~\mbox{GeV}$~\cite{tuhin24,calmet21} and $0.1~\mbox{fm}\leqslant C_V \leqslant 5~\mbox{fm}$~\cite{tuhin24,qian14}, were taken from~\cite{tuhin24}. As an illustration, we display BDM and FDM cases in Fig.~\ref{fig:eosdm-f}{\color{blue}(a)} for fixed $f$.
Notice that for the chosen set of parameters, the fermionic EOS is stiffer in comparison with the BDM one, as expected. For the sake of completeness, we also display the sound speed related to the dark matter and defined as $v^2_{s\dm}=\partial P_{\dm}/\partial\mathcal{E}_{\dm}$, see Fig.~\ref{fig:eosdm-f}{\color{blue}(b)}. From the figure, it is clear that the DM models considered here are causal, i.e., there is no superluminal equations of state in the dark sector.

Since the connection $\mathcal{E}_{\dm}=f\mathcal{E}_{\vis}$ is established, we solve the two-fluid TOV equations for a huge set of values for $f$. For each solution related to a particular $f$, we evaluate $F_{\dm}$ and select, among all the solutions, only the $(M,R)$ pairs that satisfy the particular value of $F_{\dm}$ chosen. The final result consists of a mass-radius diagram representing quark stars, each one of them admixed with a fixed fraction of dark matter.  

As a remark, we emphasize that the two-fluid approach requires a detailed analysis of the stability of the final compact quark-DM star obtained, as the reader can verify in~\cite{laura24,yunes23}. In a one fluid system, the search for stable configurations consists of investigating the response of the star to radial oscillations in a Sturm-Liouville problem for the relative radial displacement and the pressure perturbation, both quantities depending on time through $e^{i\omega t}$ with $\omega$ being the eigenfrequency. Basically, the solutions of the new differential equations that emerged from this investigation determine that stable stars are those in which $\omega^2>0$. Solutions presenting $\omega^2<0$ correspond to an exponential increase of the perturbation, leading, consequently, to the collapse of the compact star. In the case of homogeneous stars, i.e., those presenting just one phase, the location of the last stable star ($\omega=0$) coincides with the maximum mass point in the mass-radius diagram, equivalent to the point in which $\partial M/\partial\mathcal{E}^c=0$. 

\begin{figure}[!htb]
\includegraphics[scale=0.29]{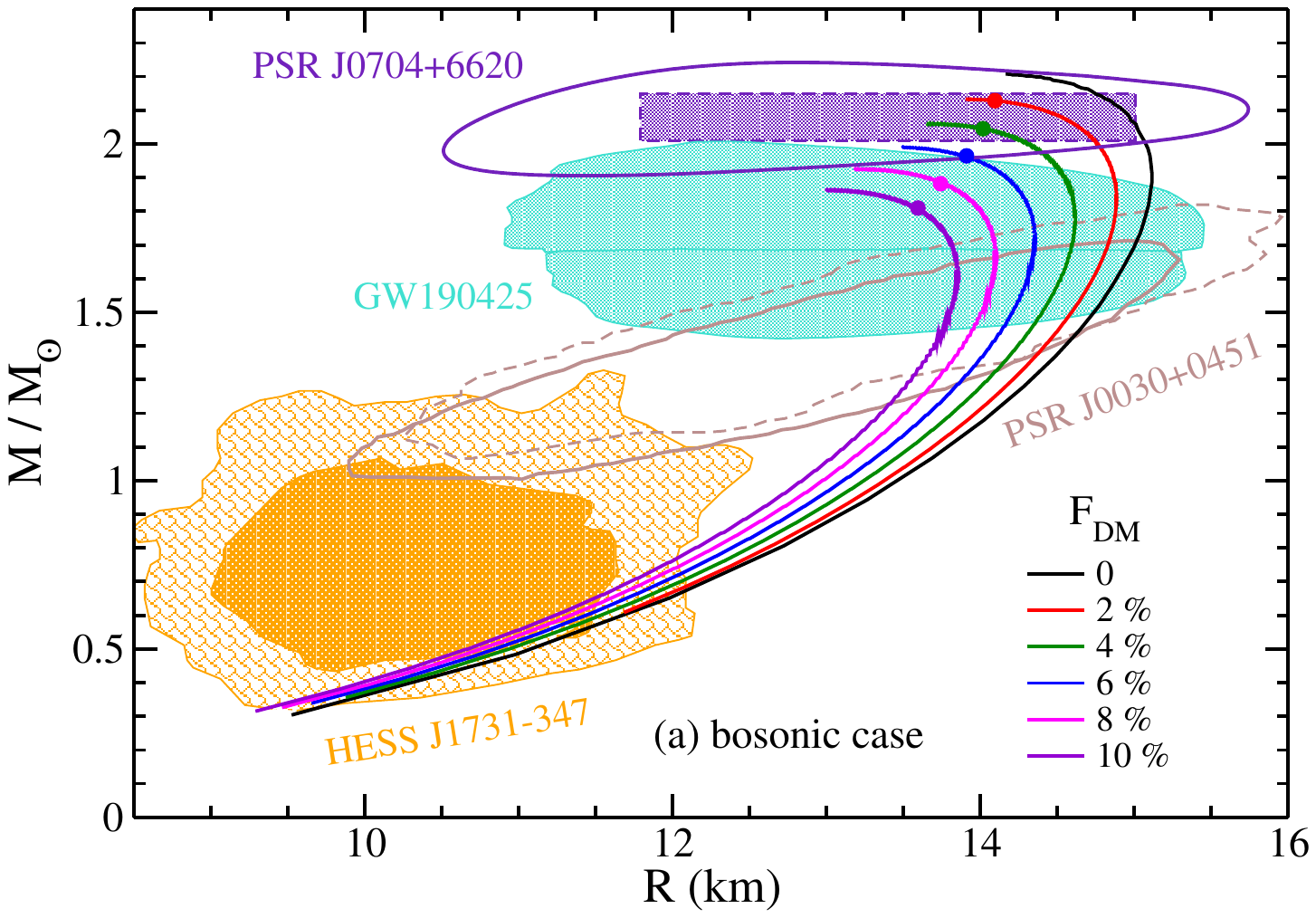}\vspace{-1cm}
\includegraphics[scale=0.29]{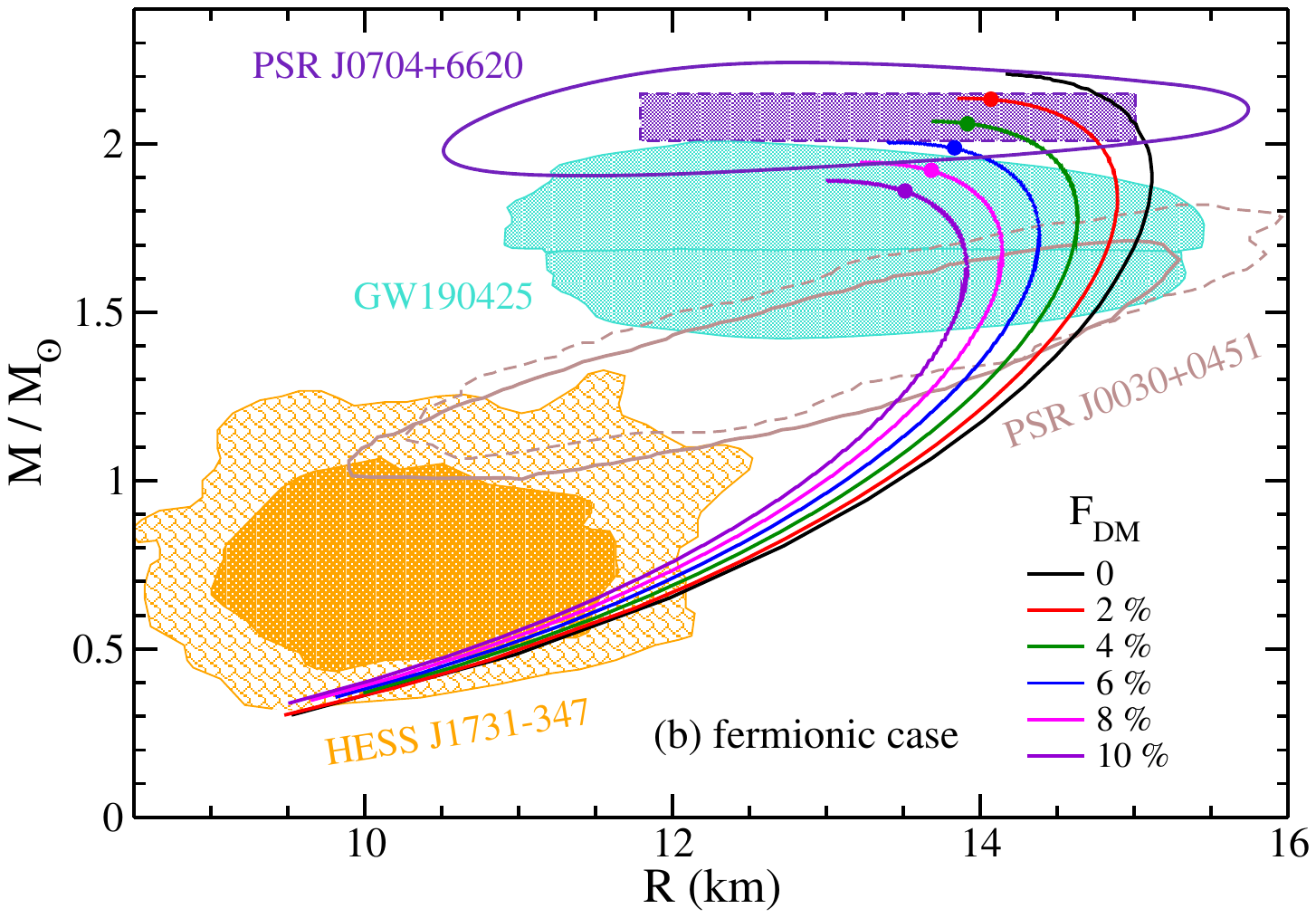}
\caption{Mass-radius relations for different fractions of dark matter in each strange stars. Results for (a)~bosonic and (b)~fermionic cases. It also plotted data related to the NICER mission concerning the pulsars PSR~J0030+0451~(dashed~\cite{Miller_2019} and solid~\cite{Riley_2019} brown contours), and PSR~J0740+6620~(solid indigo countour~\cite{Salmi_2024} and indigo rectangular band~\cite{Dittmann_2024}). We also compared our results with data from the GW190425 event~\cite{Abbott_2020-2}~(turquoise contours), and the compact object named HESS J1731-347~\cite{nat3}~(orange contours). The circles represent the last stable stars obtained from the condition given in Eq.~\eqref{condstab}.}
\label{fig:mr}
\end{figure}

In the case of the two-fluid method, the generalization determines that at the onset of unstable configurations the number of visible~($N_{\vis}$) and dark~($N_{\dm}$) particles are stationary under variations of $\mathcal{E}^c_{\vis}$ and $\mathcal{E}^c_{\dm}$, with $N_{\vis}$ and $N_{\dm}$ calculated from the differential equations given by
\begin{align}
\frac{dN_{\vis}}{dr} &= \frac{4\pi r^2\rho_{\vis}}{[1 - 2m(r)/r]^{1/2}},
\end{align}
and
\begin{align}
\frac{dN_{\dm}}{dr} &= \frac{4\pi r^2\rho_{\dm}}{[1 - 2m(r)/r]^{1/2}},
\end{align}
simultaneously coupled to the two-fluid TOV equations. As pointed out in~\cite{yunes23,laura24,pitz24,carline24}, the stability analysis is performed by investigating the diagonal version of the matrix $\partial N_i/\partial\mathcal{E}^c_j$ \mbox{($i,j$=VIS, DM)}, that comes from the equation
\begin{align}
\begin{pmatrix}
\delta N_{\vis} \\ 
\delta N_{\dm}
\end{pmatrix}
=
\begin{pmatrix}
\partial N_{\vis}/\partial \mathcal{E}^c_{\vis} & 
\partial N_{\vis}/\partial \mathcal{E}^c_{\dm} \\ 
\partial N_{\dm}/\partial \mathcal{E}^c_{\vis} & 
\partial N_{\dm}/\partial \mathcal{E}^c_{\dm}
\end{pmatrix}
\begin{pmatrix}
\delta \mathcal{E}^c_{\vis} \\ 
\delta \mathcal{E}^c_{\dm}
\end{pmatrix}
= 0.
\end{align}
The associate eigenvalues, named here as $\kappa_1$ and $\kappa_2$, are claimed to satisfy the condition 
\begin{align}
\kappa_1>0,\qquad \kappa_2>0
\label{condstab}
\end{align}
for stable stars' configurations. However, as discussed in~\cite{hong24}, there is an indication that DM models presenting central pressures in the range of $P^c_{\dm}\lesssim 5.1$~fm$^{-4}$ generate stars in a stable region. Both DM models used in this work, bosonic and fermionic ones, are in agreement with this constraint. Furthermore, the authors of~\cite{burgio1,burgio2} have used in their calculations, based on previous studies~\cite{leung12,leung22}, that the maxima of the fixed $F_{\dm}$ curves represent the last stable stars in the mass-radius diagram, as it is in the case of one fluid systems. Here we present in Fig.~\ref{fig:mr} the final mass-radius diagrams for both \mbox{quark-DM} models previously described. We also mark with circles the last stable stars ($F_{\dm}>0$) obtained from the implementation of the condition given in Eq.~\eqref{condstab}.
It is worth noticing that, as expected, for small values of $F_{\dm}$, the last stable star~($M_{ls}$) is not so different from the star with maximum mass~($M_{\max}$). The extreme case is the curve related to $F_{\dm}=0$ (no DM included) for which $M_{ls}=M_{\max}$. 

\begin{figure}[!htb]
\includegraphics[scale=0.29]{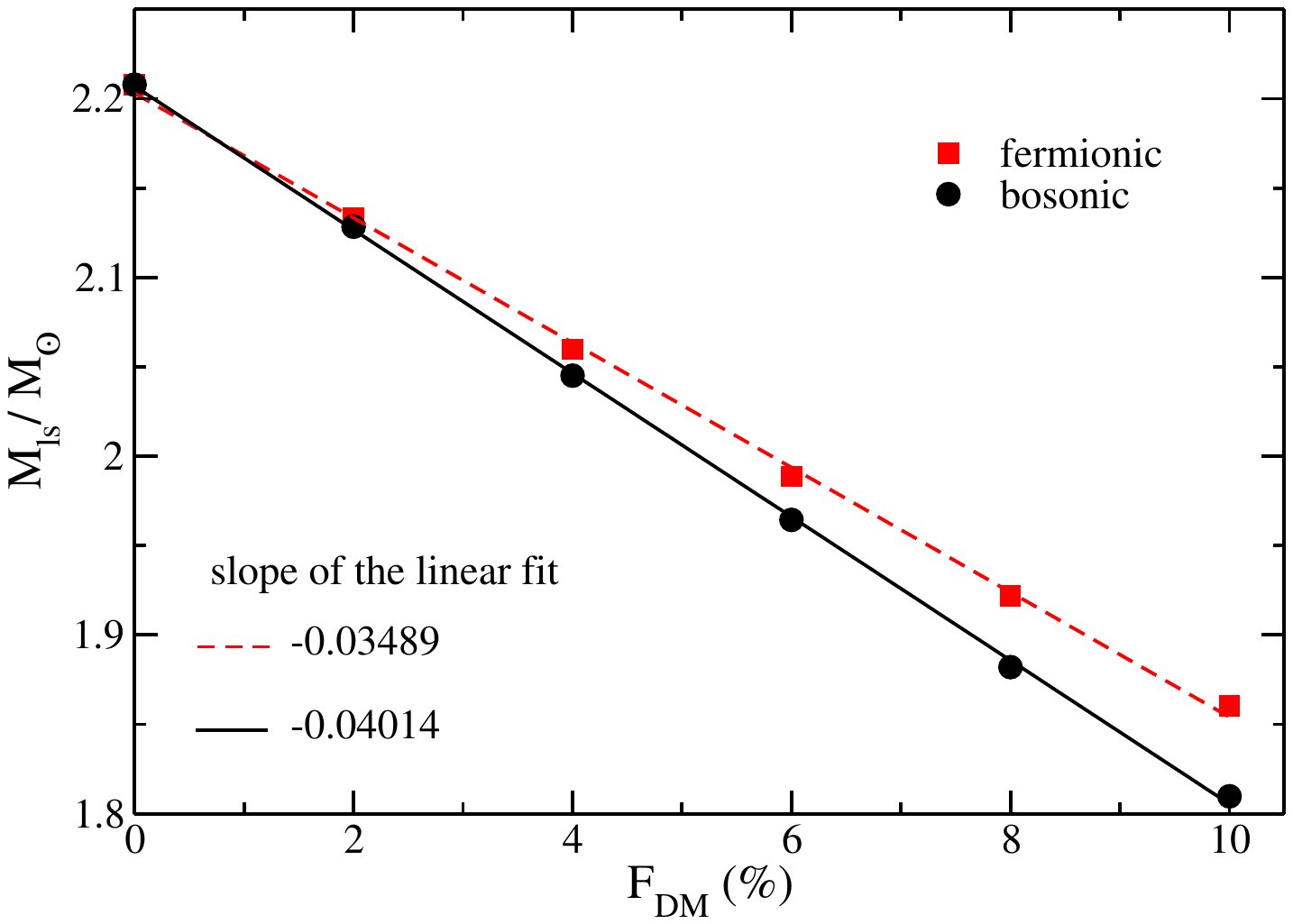}
\caption{Mass of the last stable star, $M_{ls}$, versus mass fraction for the strange-DM stars represented by circles in Fig.~\ref{fig:mr}.}
\label{fig:mmax}
\end{figure}

As another feature presented in the figure, one can verify that $M_{ls}$ of each diagram decreases as the mass fraction increases. This is a result compatible with previous studies that predict such a reduction in star configurations in which a DM core is observed~\cite{nelson19,ellis18,sagun20,leung11,sagun22}. As an illustration, we plot~$M_{ls}$ versus $F_{\dm}$ for both cases (BDM and FDM models) in Fig.~\ref{fig:mmax}.
It is worth noticing the strong linear relation between the mass of the last stable star and the dark matter content. As a consequence, one has a constant decreasing rate of $M_{ls}$ as a function of $F_{\dm}$. Such behavior is hard to predict before calculations due to the nontrivial structure of both, the two-fluid equations and the equations of state used as input, in our case, the EQP model in the visible sector, and the fermionic/bosonic models in the description of dark matter. We also furnish the slope of the linear fitting obtained, namely, $-0.04014$ ($-0.03489$) for the bosonic (fermionic) model. Notice that both are very similar. Furthermore, these values are very close to $-0.0362$, the result found in~\cite{jurgen16} in a study performed with a completely different quark model in the ordinary matter sector: the MIT bag model. For the dark sector, they have used a model made up of fermionic particles of mass $100$~GeV, a number more than $50$ times greater than the value used in the current work for the FDM model. This result may indicate a possible universality in the (linear) decreasing behavior of the maximum mass in quark models admixed with dark matter.

In Fig.~\ref{fig:mmax} we can also verify that the FDM model produces $M_{ls}$ slightly greater than the ones generated from the BDM model, a feature observed for the range of parameters investigated in our analysis. It suggests that DM equations of state are stiffer for the fermionic case with fixed dark matter content in comparison with the bosonic description of the dark sector as, indeed, we already shown in Fig.~\ref{fig:eosdm-f}. This particular figure was plotted for a fixed~$f$, but since the equations of state given as input are the same, the curves produced for a fixed value of $F_{\dm}$ are also the same.

\begin{figure}[!htb]
\includegraphics[scale=0.34]{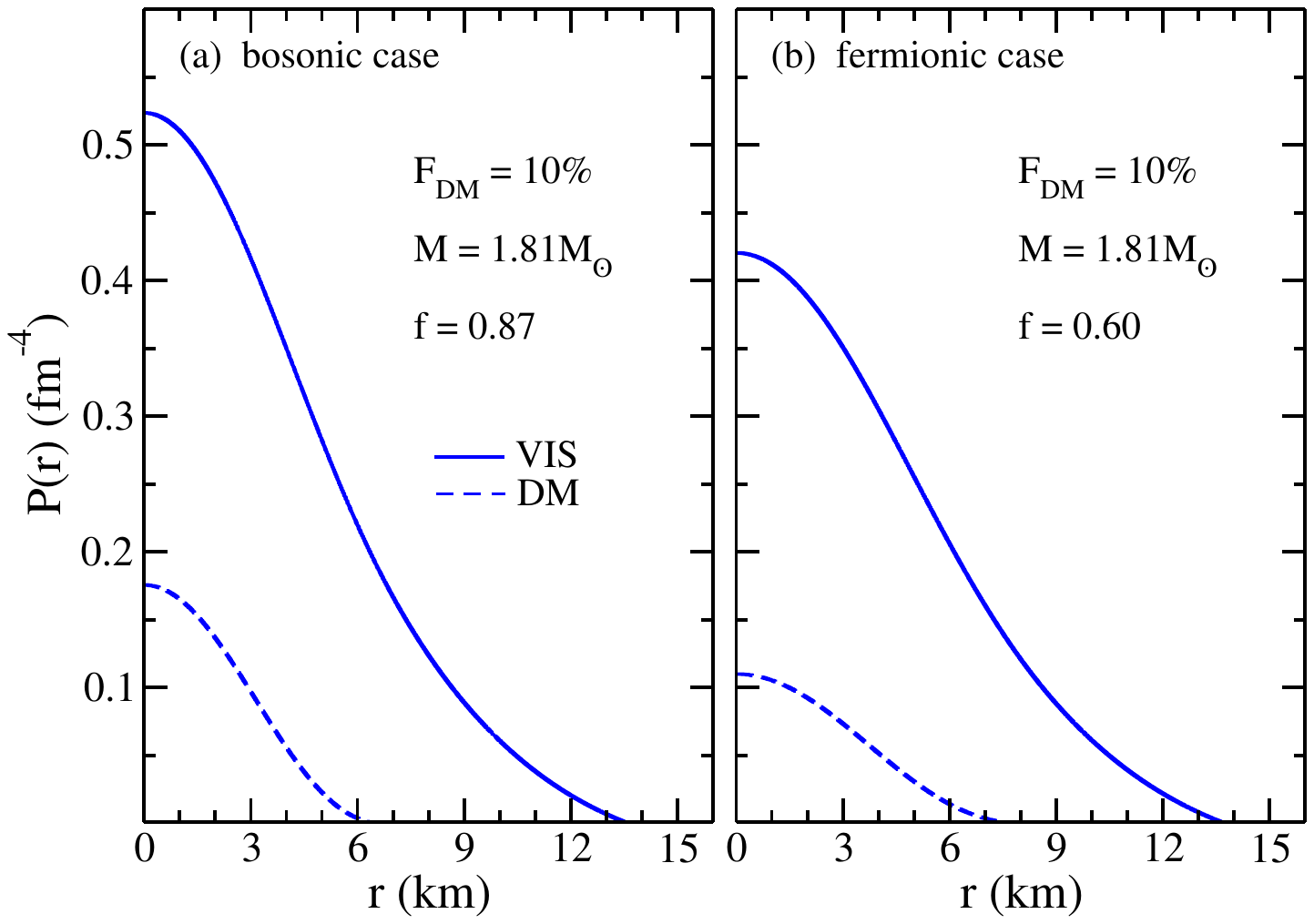}\vspace{-1.2cm}
\includegraphics[scale=0.34]{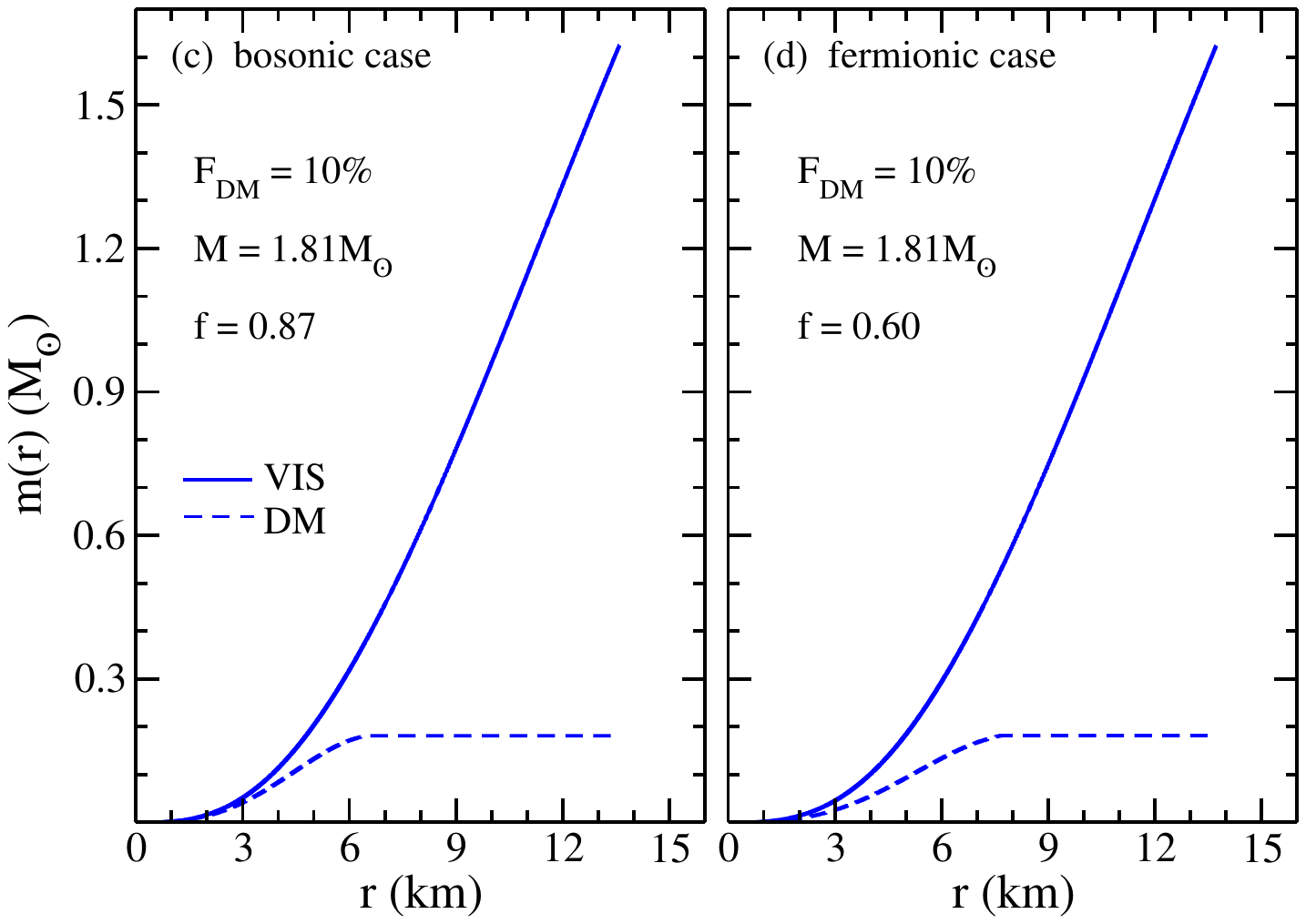}
\caption{Pressure (panels a and b) and enclosed mass (panels c and d) as functions of the radial coordinate for a mass fraction equal to $10\%$ and total stellar mass of $M=1.81M_\odot$. Full~(dashed) lines: visible~(dark) matter. For the bosonic~(fermionic) case in panel a~(b), the energy density fraction is $f=0.87$~($0.6$).}
\label{fig:profile}
\end{figure}

Still, regarding the features related to Fig.~\ref{fig:mr}, we emphasize that the difference between the entire mass-radius curves generated from the quark-FDM approach and from the quark-BDM one is very small. We assign this effect to the dominance of the effective quark model over both, bosonic and fermionic DM models used in the analysis. In order to illustrate this statement, we display in Figs.~\ref{fig:profile} the profiles of the same star constructed by using both BDM and FDM models, with $M=1.81M_\odot$ and $10\%$ of dark matter in its composition.

\begin{figure}[!htb]
\includegraphics[scale=0.34]{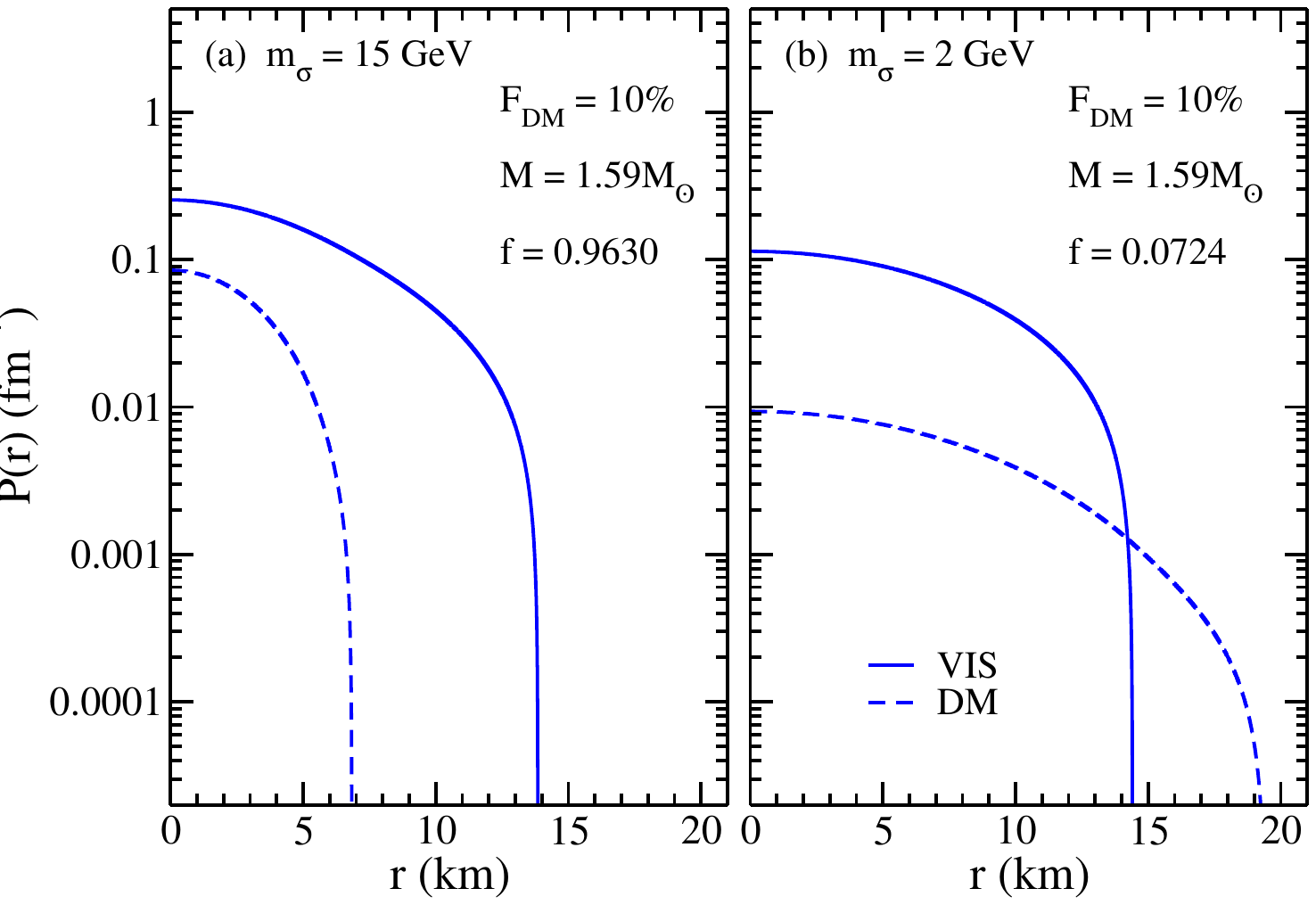}
\caption{Pressure as a function of the radial coordinate for a mass fraction equal to $10\%$ and total stellar mass of $M=1.59M_\odot$. Full~(dashed) lines: visible~(dark) matter. Strange star admixed with the bosonic dark matter with $m_\sigma=15$~GeV~(panel~a) and $m_\sigma=2$~GeV~(panel~b).}
\label{fig:pr-halo}
\end{figure}

Notice from this example that the radius of the star is $R=R_{\vis}\simeq 13.6$~($13.7$)~km, since $R_{\vis}>R_{\dm}$ in this case, by using the bosonic (fermionic) model in the dark sector. The main modification caused by the interchange of the DM model is the central pressure (at $r=0$) in both sectors. Nevertheless, $R_{\vis}$ is still practically equal for both models and determines the radius of the final star, according to the criterion used to define this quantity ($R$ corresponds to the outermost radius between $R_{\vis}$ and $R_{\dm}$). However, the situation is not the same for stars in which a DM halo is formed around it. We can verify such an effect for the case of $M=1.59M_\odot$ and $F_{\dm}=10\%$, depicted in Fig.~\ref{fig:pr-halo}, for the strange star constructed with the inclusion of bosonic dark matter in particular.
In this figure, we present two parametrizations of the BDM model, specifically with different values of the dark boson mass, namely, $m_\sigma=15$~GeV (standard value previously used) and $m_\sigma=2$~GeV. Notice that the decrease of the dark particle mass induces the situation of $R_{\dm}>R_{\vis}$ producing, consequently, the formation of a DM halo. In this case, the radius of the final star changes from $R=R_{\vis}\simeq 14$~km to $R=R_{\dm}\simeq 19$~km. Since fermionic and bosonic models generate similar mass-radius diagrams for $m_\sigma=15$~GeV and $m_\chi=1.9$~GeV, the decrease of $m_\sigma$ will produce a more pronounced difference between the diagrams obtained from FDM model with $m_\chi=1.9$~GeV in comparison with the ones determined from BDM model with $m_\sigma=2$~GeV, due to the emergence of DM halos in the latter system. These features are compatible with studies presented, for instance, in~\cite{sagun22} where the authors used the two-fluid approach with bosonic DM and in~\cite{routaray2023} where a fermionic DM model is applied in the description of stellar matter. In these works, the DM halo formation is observed when the dark particles (bosons or fermions) are decreased. In particular, since they treat neutron stars, hadronic models are used in the visible matter sector. In our work, on the other hand, strange stars are studied but the same effect is exhibited as well. 

We also emphasize some astrophysical consequences of the findings presented in our work. Recently, some studies pointed out the impact of DM on the cooling process of neutron stars, as one can see in~\cite{fibger2024,constanca2025} for instance. The authors of~\cite{fibger2024} conclude that, despite being thermally inert, DM indirectly impacts the thermal evolution of neutron stars. In that case, the one-fluid approach is adopted, and the study shows that admixed DM systems with higher values of dark Fermi momentum (dark matter content) cause an acceleration of thermal evolution, namely, a more rapid cooling of the star. In Ref.~\cite{constanca2025}, on the other hand, the two-fluid formalism is used, but the same qualitative feature is found, i.e., an increase in the DM fraction $F_{\dm}$ induces more efficient and rapid cooling due to the earlier onset of the Direct Urca process. The effects of DM on the cooling of strange stars (two-fluid formalism) could also be studied in future work to verify whether the previously observed features also arise in these compact objects. 

In particular, it could help determine whether the HESS J1731-347 measurement~\cite{nat3} can be used as a possible signature of strange stars, as proposed in~\cite{ishfaq23,diclemente2024,chu23,chu_prd23,rodrigo2023,oikonomou23,harish2023,horvath2023}. Notice that the strange stars presented here are compatible with these data, regardless of the DM content. However, unpaired quark matter could lead to difficulties in reproducing the surface temperature of this particular compact object, according to~\cite{rodrigo2023}.

Last but not least, we remark the compatibility of our results with the recent observational astrophysical data provided by the NICER mission regarding the millisecond pulsars PSR~J0030+0451~\citep{Riley_2019,Miller_2019} and PSR~J0740+6620, and with data from the gravitational wave event GW190425~\citep{Abbott_2020-2} analyzed by LIGO and Virgo Collaboration. With regard to the PSR~J0030+0451 pulsar, of mass, corresponding to $M = (2.08\pm0.07) M_\odot$~\cite{Fonseca_2021}, we have included updated data analyzed by two different groups. The first one used data from 3.6 years of observation (from 2018 September to 2022 April)~\cite{Salmi_2024}, and the second one, 
performed in Ref.~\cite{Dittmann_2024}, incorporated previous NICER and XMM-Newton observations along with additional NICER data of about $1.1$~Ms of observation in comparison with previous studies~\cite{Miller_2021,Riley_2021}. In general, our results point out possible scenarios in which strange stars can admit dark matter in their interior.

\section{Summary and concluding remarks}
\label{sec:summary}

In this work, we have investigated possible scenarios regarding the existence of strange quark stars with dark matter in their composition. The description of ordinary strongly interacting matter was carried out using the so-called equiparticle (EQP) model~\cite{Xia:2014,Backes:2020}, in which the constituent quark masses ($u$, $d$, $s$ quarks) are influenced by the medium through appropriate baryonic density-dependent functions. The parameters of these functions were chosen based on the analysis of the stellar matter stability window, ensuring parametrizations with lower energy per baryon compared to iron nuclei. Similar studies were conducted in~\cite{hong24,jurgen16}, but they relied on versions of the MIT bag model to describe the ``visible'' sector of the system. An improved version of the EQP model was proposed in~\cite{Marzola:2022a,Marzola:2023}, incorporating the Polyakov loop to account for the gluonic dynamics of strong interactions. As a result, the new model was found to be capable of describing the first-order phase transition of quark matter from a confined to a deconfined phase in the zero-temperature regime.

The dark matter component of the system was studied from two different models, namely, a fermionic and a bosonic one. The former is described by a Lagrangian density in which the dark spinor experiences a repulsive interaction mediated by a dark vector field. The full equations of state only depend on two free parameters: the dark fermion mass~($m_\chi$), and the ratio between the strength of the vector interaction~($g_V$) and the vector field mass~($m_V$). We consider these quantities within the ranges $0.5~\mbox{GeV} \leqslant m_\chi \leqslant 4.5~\mbox{GeV}$~\cite{tuhin24} and $0.1~\mbox{fm}\leqslant C_V=g_V/m_V \leqslant 5~\mbox{fm}$~\cite{tuhin24}. With regard to the bosonic model, we consider a version where we neglect interactions between the dark boson and the quarks. As in the fermionic case, we also take into account a vector interaction, in this case crucial in order to avoid the collapse of the resulting star due to the absence of degeneracy pressure. The final equations for energy density and pressure are derived and, once again, two free parameters are needed to completely determine the model: the dark scalar mass~($m_\sigma$) and $C_{\sigma\phi}=g_\sigma/m_\sigma$, where~$g_\sigma$ is the interaction strength of the dark scalar complex field with the dark vector field. Here, we adopt values consistent with
$10^{-2}~\mbox{MeV} \leqslant m_\sigma \leqslant 10^8~\mbox{MeV}$ and 
$10^{-2}~\mbox{MeV}^{-1} \leqslant C_{\sigma\phi} \leqslant 10^3~\mbox{MeV}^{-1}$~\cite{watts23}.

Since both luminous and dark matter models are defined, we allow the two sectors to interact solely through gravity, following a two-fluid approach. Our results indicate that strange stars admixed with dark matter, with a fixed dark matter fraction, constitute a possible scenario consistent with astrophysical data from massive millisecond pulsars, such as PSR~J0030+0451~\cite{Riley_2019,Miller_2019} and PSR~J0740+6620~\cite{Riley_2021,Miller_2021}, as well as observations from NASA’s Neutron Star Interior Composition Explorer~(NICER) x-ray telescope and the compact star HESS~J1731-347~\cite{nat3}.

The fourth and fifth observation runs (O4 and O5) of the LIGO-Virgo-KAGRA Collaboration are expected to significantly enhance our understanding of compact stars (CS) by substantially increasing the number of detectable gravitational wave (GW) events, with the potential detection of electromagnetic counterparts~\cite{Morsony2024}. These observations will offer unprecedented opportunities to refine the mass-radius relationship of such objects, providing critical insights into the equation of state of ultra-dense matter. Moreover, the presence of dark matter within these systems could affect GW signals, particularly by modifying post-merger waveform modes~\cite{karydas2024,Bezares2019}. The next generation of ground-based interferometers, such as the Einstein Telescope, will further improve the precision of these measurements, enabling the identification of potential dark matter signatures in binary mergers~\cite{Koehn2024}.

Complementing the analysis of these GW events, observational programs such as ATHENA~\cite{athena}, STROBE-X~\cite{strobex}, and the Enhanced X-ray Timing and Polarimetry Mission~(eXTP)~\cite{zand2019} will play an important role in advancing our understanding of CS and their potential DM content. With its high effective area and microsecond timing resolution \cite{Santangelo2024}, eXTP, for instance, will enable precise measurements of pulsations and quasiperiodic oscillations providing critical data to determine the mass-radius relationship. Furthermore, eXTP sensitivity to CS surface temperatures will allow for rigorous tests of theoretical predictions regarding DM accumulation inside CS. It is expected that the combination of these data will provide us with enough elements to determine tighter constraints on the mass-radius relationship of CS and probe the presence and the amount of DM accumulated inside these compact objects.

\section*{Acknowledgements}

\noindent
This work is part of the project INCT-FNA proc. No. 464898/2014-5. It is also supported by Conselho Nacional de Desenvolvimento Cient\'ifico e Tecnol\'ogico (CNPq) under Grants No. 307255/2023-9 (O.~L.), 401565/2023-8 (Universal, O.~L.), and doctorate scholarships No. 400879/2019-0 (I.~M), and 140324/2024-0 (A.~F.~C). E.~H.~R. was supported by the doctorate scholarship from Coordena\c c\~ao de Aperfei\c coamento de Pessoal de N\'ivel Superior (CAPES). O.~L. acknowledges M.~Dutra for stimulating discussions and valuable comments.

\bibliographystyle{apsrev4-2}
\bibliography{references-revised4}

\end{document}